\documentclass[preprint2]{aastex}
\include{epsf}
\newcommand{\be}{\begin{equation}}
\newcommand{\ee}{\end{equation}}
\newcommand{\ba}{\begin{eqnarray}}
\newcommand{\ea}{\end{eqnarray}}

\def\simless{\mathbin{\lower 3pt\hbox
   {$\rlap{\raise 5pt\hbox{$\char'074$}}\mathchar"7218$}}}
\def\simgreat{\mathbin{\lower 3pt\hbox
   {$\rlap{\raise 5pt\hbox{$\char'076$}}\mathchar"7218$}}}   % > or of order

\shorttitle{Reconstruction of Einstein Ring LBG\,J2135$-$0102}
\shortauthors{Dye et al.}

\begin{document}

%% one line. However, use \\ to force a line break.

\title{Separation of the visible and dark matter in the 
Einstein ring LBG\,J213512.73$-$010143}

%% from AASTeX v4.0. \email can be used to mark an email address
%% As in the title, \\ forces line breaks.

\author{ Simon Dye\altaffilmark{1},
Ian Smail\altaffilmark{2}, A.\,M.\ Swinbank\altaffilmark{2},
H. Ebeling\altaffilmark{3}, A. C. Edge\altaffilmark{2}
}
\affil{}
\altaffiltext{1}{Cardiff University, School of Physics \& Astronomy, 
Queens Buildings, The Parade, Cardiff, CF24 3AA, U.K.}
\altaffiltext{2}{Institute for Computational Cosmology, Durham University, 
South Road, Durham, DH1 3LE, U.K.}
\altaffiltext{3}{Institute for Astronomy, 2680 Woodlawn Drive, 
Honolulu, HI 96822, U.S.A.}

\begin{abstract}
We model the mass distribution in the recently discovered Einstein
ring LBG\,J213512.73-010143 (the `Cosmic Eye') using archival {\sl Hubble
Space Telescope} imaging.  We reconstruct the mass density profile of
the $z=0.73$ lens and the surface brightness distribution of the
$z=3.07$ source and find that the observed ring is best fit with a
dual-component lens model consisting of a baryonic Sersic component
nested within a dark matter halo. The dark matter halo has an inner
slope of $1.42^{+0.24}_{-0.22}$, consistent with CDM simulations after
allowing for baryon contraction.  The baryonic component has a
mass-to-light ratio of $1.71^{+0.28}_{-0.38}$ M$_{\odot}$/L$_{B\odot}$
which when evolved to the present day is in agreement with local
ellipticals. Within the Einstein radius of $0.77''$ (5.6 kpc), the
baryons account for $(46\pm 11)$\% of the total lens mass.  External
shear from a nearby foreground cluster is accurately predicted by the
model. The reconstructed surface brightness distribution in the source
plane clearly shows two peaks. Through a generalisation of our lens
inversion method, we conclude that the redshifts of both peaks are
consistent with each other, suggesting that we are seeing structure
within a single galaxy.
\end{abstract}


\section{Introduction}

The measurement of galaxy mass profiles has proven a powerful
observational probe for testing models of structure formation.  In
recent years, the slope of the inner mass profile has become a subject
of much contention with the results of cold dark matter (CDM)
simulations being discrepant with observations. At the present time,
this discrepancy persists.

\citet{nfw96} first proposed an analytic approximation to describe the
mass density profiles of halos in their CDM simulations. At a radius
much smaller than a particular scale radius, the mass density of this
so called `NFW' profile scales as $\rho(r) \propto r^{-1}$. Later
simulations by \citet{moore98,moore99} indicated a steeper inner
slope. This gave rise to the `generalised NFW' (gNFW) profile which at
radii smaller than the scale radius follows $\rho(r) \propto
r^{-\alpha}$ with values of $\alpha$ around 1.4 to 1.5.  The most
recent simulations boasting a significantly higher resolution now
converge on a slope somewhere in the range $1.0\simless \alpha
\simless 1.2$ \citep{navarro04,diemand05}.

These results appear to strongly disagree with measurements of the
inner slope from dynamical studies. Several groups measuring rotation
curves of low surface brightness galaxies (LSBs; these are believed to
have a high dark matter fraction and therefore be minimally affected by
baryons) find a range of slopes; $0\simless \alpha \simless 1$
\citep{deblok01,deblok02,swaters03,spekkens05}. \citet{hayashi04}
purported that this discrepancy could be reconciled by directly
comparing against rotation curves of simulated halos. However, this
was strongly rejected by \citet{deblok05} who found that only one
quarter of the 51 galaxies in the study of \citet{hayashi04} were
consistent with CDM. In the latest episode of this ongoing debate,
\citet{hayashi06} claim that non-circular motions in simulated CDM
halos arising from a triaxial potential can explain the range of
measured rotation curves seen in LSBs. Clearly dynamical measurements,
whilst potentially very powerful, have many complexities preventing a
straight-forward interpretation.

Gravitational lensing has for some time now provided an attractive
alternative means of measuring mass profiles without the difficulties
associated with dynamical techniques.  Primarily, this is motivated by
the simple fact that the deflection angle of a photon passing a
massive object is independent of the dynamical state of the mass
within the object. Strong lensing systems with multiple images of a
background source can constrain the radial profile of the projected
mass density of the lens by searching for the best fit to the observed
image positions \citep[for example, see the review by][]{schneider06}.
This technique was enhanced by \citet{sand02} to incorporate extra
constraints from the observed velocity dispersion profile of the lens
and has since been applied to a number of systems
\citep{treu02,koopmans03,sand04}. 

\citet[][hereafter DW05]{dye05} showed how Einstein ring systems,
i.e., strong lens systems where an extended source is imaged into a
complete or near-complete ring, can place stronger constraints on the
mass profile of the lens than systems with multiple point-like
images. This work used the semi-linear method of \citet{warren03}
which has also been used by several other studies to date
\citep{treu04,treu06,koopmans06}.  A Bayesian version of the
semi-linear method was developed by \citet{suyu06} and later enhanced
by \citet{barnabe07} with the inclusion of linear constraints from
stellar velocity moments.

In this paper, we follow the procedure of DW05. We apply the
semi-linear method to reconstruct the lens mass profile and source
surface brightness image of the extraordinary Einstein ring system
LBG\,J2135-0102 recently discovered by \citet[][see also Coppin et al.
2007]{smail07}. Due to the lens amplification, this system is of
particular interest as it represents one of the brightest examples of
Lyman break galaxies known.

We compare the fit given by five different lens models. With
our most general model which comprises a gNFW halo that hosts a mass
component following the lens galaxy light, we separate the baryonic
and dark matter contributions to the projected lens mass within the
Einstein radius.  We also describe a simple extension of the
semi-linear method enabling reconstruction of sources in multiple
planes at different redshifts.  Our analysis adds to the growing list
of Einstein ring systems that are now beginning to provide strong
constraints on CDM simulations.

The layout of this paper is as follows. In the following section we
briefly describe the data. Section \ref{sec_sl_method} outlines the
semi-linear method and our lens models. We present the results of our
source and lens reconstruction in Section \ref{sec_results}. Finally,
we conclude with a summary and discussion in Section
\ref{sec_summary}. Throughout this paper, we assume the following
cosmological parameters; ${\rm H}_0=70\,{\rm km\,s}^{-1}\,{\rm
Mpc}^{-1}$, $\Omega_m=0.3$, $\Omega_{\Lambda}=0.7$.

\section{Data}
\label{sec_data}

\begin{figure}
\epsfxsize=7.0cm
{\hfill
\epsfbox{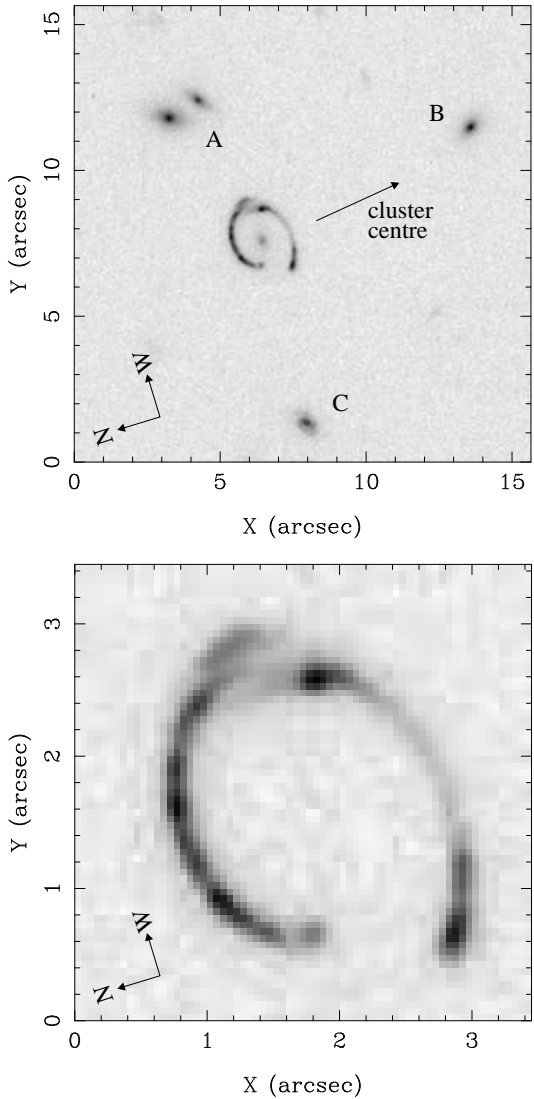}
\hfill}
\epsfverbosetrue
\caption{The ACS Wide Field Camera data observed in F606W. North lies at
an angle of $197.4^{\circ}$ measured counter-clockwise from the
positive $x$-axis.  {\em Top}: The Einstein ring and lens galaxy
showing the nearby likely cluster members labeled A, B
and C. The cluster MACS J2135.2-0102 lies $75''$ away to the south as
indicated.  {\em Bottom}: The ring after subtraction of the lens
galaxy satisfactorily fit with a Sersic profile.}
\label{obs_data}
\end{figure}

The image data were acquired as part of the Snapshot programme GO
\#10491 (PI: H. Ebeling) on May 8th 2006 using the {\sl Hubble Space
Telescope} ({\sl HST}) Advanced Camera for Surveys (ACS). Imaging of
the Einstein ring itself was completely serendipitous, the intended
target being the cluster MACS J2135.2-0102 lying approximately $75''$
away to the south. The top panel of Figure \ref{obs_data} shows a
section of the reduced ACS image with the Einstein ring and lens
galaxy just to the left of centre.  The cluster centre lies at
$24^{\circ}$ in a counter-clockwise direction measured from the +ve
$x$-axis.  The nearby galaxies labeled A, B and C are likely cluster
members. We consider the perturbative effect from these nearby
galaxies and the large scale shear from the cluster itself on the lens
solution in Section \ref{sec_lens_recon}.  

The image is a stack of three dithered 400s exposures taken with the
ACS Wide Field Camera (WFC) in the F606W filter. The stack was
produced using the STScI {\sc MULTIDRIZZLE} package (V.2.7) thereby
correcting for the geometric shear introduced by the WFC and giving a
resulting pixel scale of $0.05''$.  The reader is referred to
\citet{smail07} for more details.

Keck spectroscopy taken on June 30th 2006 and July 24th 2006
unambiguously identifies the lensed source as a Lyman break galaxy
lying at a redshift of $z=3.07$.  Similarly, the lens galaxy spectrum
exhibits several strong absorption and emission features placing it at
a redshift of $z=0.73$. The total magnitude of the lens galaxy in the
F606W filter is $R_{F606W}=22.34\pm0.15$ \citep{smail07}. In addition
to the ACS imaging data, \citet{smail07} obtained a $K'$ image using
the Near Infra-Red Camera at Keck II on July 4th 2006. The
($R_{F606W}-K'$)=2.8 colour of the lens and the spectral continuum is
consistent with that of an early type spiral or S0 galaxy.

Before reconstructing the source surface brightness distribution, any
contribution of flux from the lens galaxy to the ring flux must be
removed. Although the lens contribution in this case is unlikely to be
of concern, we carried out this exercise anyway to establish the
visible morphology of the lens for our modelling later.  An added complication
in this case is that the lens galaxy light is sheared by the foreground
cluster and this must be accounted for in the lens model (Section
\ref{sec_model_proc}). We  fitted an elliptical Sersic 
profile \citep{sersic68} of the form
\be
\label{eq_sersic}
{\rm L}={\rm L}_{1/2}\exp\{-B(n)[(r/r_s)^{1/n}-1]\} 
\ee 
to the lens galaxy having first masked out pixels within an elliptical
annulus containing the ring (see Figure
\ref{recon_source_montage}). For the fitting of this profile and for
the modelling later, we created a noise map based on the CCD gain,
read noise and photon count. The parameters ${\rm L}_{1/2}$, $n$ and
$r_s$ were allowed to vary in the fit as well as the elongation (i.e.,
major axis divided by minor axis), $e$, and the orientation (angle in
degrees of the major axis to the positive image $x$-axis in a
counter-clockwise sense), $\theta$. In the fitting, we convolved each
trial surface brightness profile with the WFC point spread function
(PSF) modelled by the {\sl TinyTim} package \citep{krist04} for the F606W
filter\footnote{The {\sl TinyTim} PSF includes CCD pixel charge smearing 
and geometric distortion of the ACS image plane.}.
The best fit was achieved with the parameters
listed in Table \ref{tab_sersic} which give an acceptable $\chi^2$.

\begin{table}
\centering
\small
\begin{tabular}{cc}
\hline
Parameter & Minimised Value \\
\hline
${\rm L}_{1/2}$ & $(3.14 \pm 0.25)\times 10^9 \,{\rm L}_{B\odot}/ \Box''$\\
$n$ & $2.90\pm0.12$ \\
$r_s$ & $0.57''\pm0.09''$ \\
$\theta$ & $110.1^{\circ}\pm2.6^{\circ}$ \\
$e$ & $1.38\pm0.07$ \\
\hline
\end{tabular}
\normalsize
\caption{Parameters of the Sersic profile fit to the lens galaxy.
Reading from top to bottom, these are; the surface brightness at the
scale radius (${\rm L}_{1/2}$), the Sersic exponent ($n$), the scale
radius ($r_s$), the orientation of the major axis to the positive
image $x$-axis in a counter-clockwise sense ($\theta$) and the
elongation, i.e. major axis divided by minor axis ($e$).}
\label{tab_sersic}
\end{table}

We work in terms of the rest-frame $B$ band luminosity in this paper
for ease of comparison with other studies. The conversion from F606W
flux (approximately rest-frame $U$) to $B$ band luminosity density was
made using a k-correction and a colour correction (see Section
\ref{sec_model}).  The fitted profile was subtracted from the unmasked
image and the result is shown in the lower panel of Figure
\ref{obs_data}.

\section{Method of Analysis}
\label{sec_sl_method}

\subsection{Semi-linear inversion}

For a full description of the semi-linear inversion method, the reader
is referred to \citet{warren03} and DW05. We provide a very
brief outline here for completeness.

The inversion assumes the source plane and the image plane are
pixelised. The manner in which the source plane is pixelised is not
restricted, allowing the `adaptive gridding' described in DW05 whereby
smaller pixels are concentrated in regions where there are stronger
constraints (see next section).  

For a given lens model parameterisation, a PSF-smeared image is
computed for every source pixel. All images are created using unit
surface brightness source pixels. The problem of finding the factor
required to scale each image such that their co-addition best fits the
observed image is a linear one. These scale factors are the best fit
surface brightnesses of the source pixels for the given lens
model. The solution is written as 
\be
\label{eq_sl}
\underline{s} = ({\rm \underline{F}} +
\lambda {\rm \underline{H}})^{-1} \underline{c} \, ,
\ee
where the square matrix ${\rm \underline{F}}$ and the vector
${\rm \underline{c}}$ have the elements
\be
{\rm F}_{ik}=\sum_j f_{ij}f_{kj}/\sigma_j^2 \, , \,\,\,
c_i=\sum_j f_{ij}d_j/\sigma_j^2
\ee
and $\underline{s}$ 
is a vector containing the best fit source pixel surface brightnesses.
Here, $d_j$ is the observed flux in image pixel $j$, $\sigma_j$ its
error and $f_{ij}$ is the flux in pixel $j$ of the image of source
pixel $i$ for the current lens model.
The solution is regularised by the square regularisation
matrix ${\rm \underline{H}}$, scaled by the regularisation weight
$\lambda$ \citep[see][and Warren \& Dye 2003]{press01}. 
The standard errors of the reconstructed source pixels are given
by the diagonal terms of the covariance matrix which is just:
\be
\label{eq_cov_matrix}
({\rm \underline{F}} + \lambda {\rm \underline{H}})^{-1}\, .
\ee

This linear inversion is nested inside an outer loop that varies the
non-linear lens model parameters. For each trial set of lens
parameters, the image of the reconstructed source is compared to the
observed image and the $\chi^2$ computed within an elliptical mask
containing the ring. The mask is designed to ensure that it includes
the image of the entire source plane, with minimal extraneous
sky. This means that only significant image pixels are used in the
fit, making $\chi^2$ more sensitive to the model parameters.

The regularisation scheme is chosen to best suit the 
reconstructed source surface brightness image. \citet{warren03}
showed using simulations that linear regularisation works well
for this application. However, as we discuss in the next
section, regularisation biases the solution between different
lens models and so we set $\lambda=0$ and use an adaptive
source grid when searching for the best fit lens parameters.

In DW05, we used a downhill simplex method to find the set of lens
parameters that minimises $\chi^2$. However, the $\chi^2$ surface is
pitted with local minima and noisy. \citet{warren03} described how
this noise can be reduced by sub-gridding the image pixels when
computing the source pixel images. In this work, we therefore divide
image pixels into $16 \times 16$ sub-pixels. To prevent becoming
trapped in local minima, we use a simulated annealing downhill simplex
minimisation algorithm. We find that a slow exponentially cooled
temperature with a half-life of $\sim 50$ iterations works extremely
well in finding the global minimum.

\subsubsection{Multiple source planes}

As we show in Section \ref{sec_src_recon}, the reconstructed
source surface brightness distribution of LBG J2135-0102 has two
distinct peaks. An interesting question to ask is therefore
whether these peaks are two separate objects at different redshifts
or, given their small angular separation, two flux
peaks within the same object. The ring spectra presented in
\citet{smail07} show no evidence of a different secondary source redshift
although the fainter source would be dominated by the
brighter and much more highly magnified primary source
\cite[see also the discussion of the CO emission from this system
in][]{coppin07}. We explain here how the semi-linear inversion method
can be modified in a simple way to place constraints on this problem.

The inversion described above is completely general in terms of
the geometry of each source pixel and the lens model. Qualitatively,
this means that each source pixel image can be created assuming a
different lens configuration.  The practical use of this is not in
changing the lens mass profile, but the lens {\em geometry} so that
different source pixels can be assigned different redshifts.  In the
case of two source planes, this quantitatively means that the vector
$\underline{s}$ in equation (\ref{eq_sl}) holds two groups of source
pixels, $s_1 \rightarrow s_n$ and $s_{n+1} \rightarrow s_{n+m}$.  The
matrices ${\rm \underline{F}}$ and ${\rm \underline{H}}$ and the
vector $\underline{c}$ are computed in exactly the same way as
described in \citet{warren03}, except the lens configuration for the
first $n$ pixels uses a different source redshift to that for the
remaining $m$ pixels.

In this paper, we carry out the usual reconstruction with a single
source plane, then a separate analysis with two source planes.  For
the second analysis, we define two source planes; plane 1 containing
pixels that belong to the brightest and more highly magnified source
and plane 2 containing pixels of the fainter source as Figure
\ref{recon_source_montage} shows.  To investigate whether the redshift
of source 2 is consistent with that of source 1, we hold plane 1 fixed
at the measured spectroscopic redshift $z=3.07$ and allow the redshift
of plane 2 to vary. This introduces an extra non-linear lens model
parameter (see Section \ref{sec_model}). In the minimisation, all lens
parameters including this additional parameter are allowed to vary.

\subsection{Adaptive source plane grid}
\label{sec_adaptive_gridding}

We adopt the gridding algorithm of DW05 in the current paper which
creates smaller source plane pixels where the magnification is higher.
The gridding algorithm starts with a regular mesh of large pixels. For
a given lens model, the average magnification $\mu_i$ of every source
pixel $i$ is computed. Those pixels that meet the criterion $\mu_i \, r_i
\geq s$ are then split into quarters, where $r_i$ is the ratio of the
area of pixel $i$ to the area of an image pixel and $s$ is the
'splitting factor'. Having finished the initial loop through all
pixels, the process is repeated for the sub-pixels, then for the
sub-sub-pixels and so on until all pixels satisfy the splitting
criterion. The entire procedure (which takes a fraction of a second on
a modern desktop computer) is repeated every time the lens model
changes.

The splitting factor is set empirically by measuring the improvement
in the fit brought about by the splitting for different values of $s$
with a fixed lens model close to the best solution. The splitting
factor is successively reduced until no significant improvement is
obtained. For our unregularised reconstruction, we find a splitting
factor of $s=9$ satisfies this condition. By initialising the adaptive
gridding with a source plane of regular $0.1'' \times 0.1''$ pixels,
this splitting factor ensures that images of split source pixels do
not exceed the Nyquist sampling resolution set by the image PSF (see DW05
for more details on how $s$ is set and how the initial grid size is
chosen). This is important since it keeps degeneracies between source
pixels to a very low level, thereby allowing the number of degrees of
freedom to be accurately determined (see below).

The main advantage of an adaptive grid is that the resolution of the
reconstructed source can be enhanced relative to a regular grid
without the need for regularisation.  Regularisation has the effect of
smoothing the reconstructed source light profile, reducing the
effective total number of parameters and hence increasing the number
of degrees of freedom, by an amount that cannot be satisfactorily
quantified. This is especially problematic when comparing different
lens models, as a fixed regularisation weight for one model generally
does not give the same increase in number of degrees of freedom for
another. This problem was noted by \citet{kochanek04}. Adaptively
sized pixels extract maximum information from the lens image without
need of regularisation. Therefore, provided the degeneracies between
all pairs of source pixels are negligible, the number of degrees of
freedom of the fit is a well--defined number. This allows, firstly,
unambiguous assessment of whether a given model provides a
satisfactory fit to the data and secondly, unbiased comparison of
different model fits.

\citet{suyu06} and \cite{barnabe07} suggest an alternative means of
circumventing this problem with a Bayesian approach. They point out
that the Bayesian evidence enables different lens models and
regularisation weights to be compared in an unbiased way. However, we
prefer to adhere to the adaptive grid technique without regularisation
since this has the appealing characteristic that the error image of
the reconstructed source is much more uniform. To ensure that the
number of degrees of freedom we compute in our unregularised solution
is accurate, we verify that the degeneracies between all source
pixel pairs are negligible by ensuring that all off diagonal terms in
the covariance matrix given by equation (\ref{eq_cov_matrix}) are
negligible.

Although regularisation is not used when quantifying lens model fits,
we use it in this paper solely for the aesthetic purpose of obtaining
a higher resolution source. In DW05, we used zeroth order
regularisation to simplify the implementation. However, this is an
over-simplistic type of regularisation since it poorly represents real
astronomical sources. A better scheme is linear regularisation where
pixels that differ more strongly in surface brightness from their
neighbours are penalised more heavily. We implement a form of linear
regularisation on the adaptive grid in the current paper by
constructing a regularisation matrix that takes the difference between
a given pixel $i$ and the sum of all neighbouring pixels $j$ weighted
by $w_{ij}=(a_j/a_i)\,N\exp(-y_{ij}^2/2\sigma^2)$. Here, $a$ is the
source pixel area, $y_{ij}$ is the separation of the centres of pixels
$i$ and $j$ and $N$ is a normalisation constant set such that
$\sum_{j,j\ne i} w_{ij}=1$. We set $\sigma=0.05''$. As described in
DW05, we use a smaller splitting factor when applying regularisation.
Figure \ref{recon_source_montage} shows the adaptive grids upon which
the regularised and unregularised source is reconstructed.

\subsection{Lens model}
\label{sec_model}

We compare the fit given by five different parameterisations of the
lens mass profile.  The most general of these is a dual component
model. In this model, one component follows the lens galaxy light (the
baryons) and the other, a gNFW profile represents the dark matter
halo. The remaining four single component models are special cases of
the dual component model; 1) a singular isothermal ellipsoid (SIE), 2)
a gNFW profile, 3) a power-law model, 4) a mass-to-light (M/L) model
where the mass follows the lens galaxy light.

The dual component model is that used by DW05.  For the baryonic
component, only the $B$ band M/L ratio, $\Psi$, is allowed to vary
during the minimisation. Its morphology is fixed by the lens galaxy
surface brightness profile that would have been observed in the
absence of cluster shear. This is simply a transformation of the Sersic
profile fit to the observed surface brightness profile (see Section
\ref{sec_model_proc}).

The dark matter halo component is described by an elliptical gNFW 
profile with a volume mass density given by
\be 
\rho(r)=\rho_s (1+r/r_0)^{\alpha-3}(r/r_0)^{-\alpha}. 
\ee
The halo has seven parameters; the normalisation, $\rho_s$, the inner
slope $\alpha$, the offset of the halo centre from the baryonic
centre, $(\Delta x_h, \Delta y_h)$, the scale radius $r_0$, the
elongation $e_h$ (equal to the major:minor axis ratio) and the
orientation of the major axis to the positive image $x$-axis in a
counter-clockwise sense, $\theta_h$. As we discussed in DW05, the
scale radius is very weakly constrained in this model, hence we hold
it fixed at the value of $r_0=3.0''$ in accordance with the results of
\citet{bullock01} for a galaxy with similar total mass and redshift
properties to the lens.

We include the effect of the nearby cluster in two ways.  Firstly, we
incorporate an external shear that is allowed to vary in magnitude,
$\gamma$, and direction, $\theta_{\gamma}$ (measured counter-clockwise
from the +ve $x$-axis). The cluster convergence must also be accounted
for, as this causes magnification of the ring image. We therefore
calculate a convergence based on $\gamma$ and $\theta_{\gamma}$
assuming a SIE profile (using a NFW profile for this purpose instead
makes little difference to the best fit lens model parameters compared
to the size of their errors). Secondly, we model perturbations in the
cluster potential from the nearby cluster galaxies.  In this case,
each galaxy is modelled as a SIE with a mass given by its total
integrated light multiplied by a common $B$ band M/L, $\Psi_{cg}$.

In summary, our dual component model has a total of ten free
parameters; $\Psi$, $\rho_s$, $\alpha$, $(\Delta x_h, \Delta y_h)$,
$\theta_h$, $e_h$, $\gamma$ , $\theta_{\gamma}$ and $\Psi_{cg}$. The
single component models are obtained by holding various combinations
of these parameters fixed. For the gNFW model, we only fix $\Psi=0$
and allow all other parameters, including the cluster parameters
($\gamma$, $\theta_{\gamma}$ and $\Psi_{cg}$) to vary. Similarly, for
the power-law and SIE models, the halo scale radius, $r_0$, is held at
an arbitrarily large value having fixed $\Psi=0$ with the additional
constraint $\alpha=2$ for the SIE. The pure M/L model allows
only $\Psi$ and the cluster parameters to vary.

The deflection angles of both the Sersic and gNFW profiles must be
numerically evaluated. Also, the surface mass density of the gNFW
profile must be obtained by numerically integrating along the line of
sight. We showed in DW05 how we convert the circular Sersic and gNFW
profiles to elliptical profiles. 

The surface mass density of the Sersic profile is obtained by
multiplying the fitted light profile given in equation
(\ref{eq_sersic}) by the M/L ratio, $\Psi$. To convert the lens galaxy
flux measured in the F606W filter to a $B$ band luminosity, we
computed a k-correction and a colour correction using a standard S0
and Sa galaxy template taken from \citet{mannucci01}. These SEDs are
consistent with the $(R-K')$ colour measured by \citet{smail07} and the
observed lens morphology.  The k-correction and colour correction for
the S0 template are respectively -1.85 mag and 2.03 mag, and for the
Sa template, -1.69 mag and 1.97 mag.  We take the average k-correction
and colour correction and treat the difference in each as a 1$\sigma$
error. Similarly, for the perturbing cluster galaxies, we assumed a
standard elliptical spectrum to convert to a $B$ band luminosity.  For
the cluster redshift of $z=0.33$, the colour correction is 2.06 mag
and the k-correction -0.63 mag.

\subsubsection{Modelling procedure}
\label{sec_model_proc}

Since the nearby foreground cluster shears the light from the lens
galaxy as well as the ring image, the morphology of the baryons cannot
be simply fixed as the Sersic profile fit to the observed lens galaxy
surface brightness (see Section \ref{sec_data}).  In theory, we could
allow the baryonic morphology to vary in the minimisation. However, in
practice, the morphology proves very poorly constrained since the
baryons have only a small elongation (see Section \ref{sec_lens_recon})
and only contribute approximately one third of the total projected
mass inside the Einstein radius (including the cluster convergence).

With this in mind, we adopt the following procedure when fitting the
lens with the dual component model and pure M/L model:

\begin{itemize}

\item[1.] First fit the gNFW, power-law and SIE models to determine 
the best fit cluster shear.

\item[2.] Using the average of this best fit shear, compute the elongation 
and orientation of the Sersic profile that would have been observed in
the absence of the cluster.

\item[3.] Fit the dual component model, fixing the baryon component
with the Sersic profile parameterised in Table \ref{tab_sersic}, except
with the elongation and orientation computed in step 2.

\end{itemize}

We compute an error on the elongation and orientation estimated in step 3
by combining the errors from the original Sersic fit, as given in Table
\ref{tab_sersic}, and the error on the shear determined in step 1.  
This resulting uncertainty is incorporated into our modelling using a
Monte Carlo simulation which randomly samples the orientation and elongation
for 100 minimisations. The final errors we quote combine the spread
in minimised model parameters with the formal errors from the fit
and additionally include the colour and k-correction error for the
M/L.

Having fit each of the five lens models, we repeat the
analysis with the best fit model and a second source plane.  In this
case an additional parameter, $z_2$, is introduced for the second
source plane redshift.

\section{Results}
\label{sec_results}

\subsection{Lens reconstruction}
\label{sec_lens_recon}

Table \ref{tab_cf_models} compares the $\chi^2$ for the five models.
The best fit to the observed ring is given by the dual component
model. The pure M/L model is very strongly ruled out. The pure halo,
power-law and SIE models give a satisfactory $\chi^2$ but perform
significantly worse than the dual component model with $\Delta \chi^2
= 17.4$ for six fewer degrees of freedom for the gNFW, $\Delta \chi^2
= 16.2$ for 12 fewer degrees of freedom for the power-law and $\Delta
\chi^2 = 22.8$ for the SIE.

\begin{table}
\centering
\small
\begin{tabular}{ccccc}
\hline
Model & $\chi^2_{\rm min}$ & NDoF & N$_{\rm src}$ & N$_{\rm par}$ \\
\hline
dual component & 1912.6 & 1896 & 278 & 10 \\
SIE & 1935.4 & 1888 & 289 & 7 \\
pure gNFW & 1930.0 & 1890 & 285 & 9 \\
power-law & 1928.8 & 1884 & 291 & 9 \\
pure M/L & 2188.4 & 1920 & 260 & 4 \\
\hline
\end{tabular}
\normalsize
\caption{Comparison of the lens models in terms of 
$\chi^2$ at the minimum. The number of degrees of freedom (NDoF) for
each model is given by the number of image pixels in the elliptical
annulus ($=2184$ -- see Figure \ref{recon_source_montage}) minus the
sum of the number of source pixels in the adaptive grid (N$_{\rm
src}$) and the number of parameters in the lens model (N$_{\rm
par}$).}
\label{tab_cf_models}
\end{table}

We fitted the gNFW, power-law and SIE models first to establish the
cluster shear (see Section \ref{sec_cluster_shear}). The normalisation
of the SIE model corresponds to a central velocity dispersion of
$210\pm10$ km s$^{-1}$, in agreement with the measurement of
$\sim 230\pm30$ km s$^{-1}$ from the lens galaxy spectrum \citep{smail07}.
The slope of the gNFW model is $1.95\pm0.03$, consistent with the
slope of $2.01^{+0.02}_{-0.03}$ averaged over 15 lenses by
\citet{koopmans06}.  Similarly, the best fit slope of the power-law
model is $2.09 \pm 0.04$.

Using the average shear obtained from the gNFW and SIE models, we
distorted the Sersic profile fitted to the observed galaxy light to
give a resulting elongation of $1.14\pm0.08$ and orientation
$23.2^{\circ}\pm4.2^{\circ}$. This sheared Sersic profile was then
used to fix the morphology of the baryonic component for the pure M/L
model and the dual component model.

The best fit parameters of the dual component model are given in Table
\ref{tab_dual_params}. The baryonic component accounts for $46\pm
11$\% of the total lens mass inside the Einstein radius of $0.77''$
(5.6 kpc).  This excludes the convergence from the cluster and its
nearby perturbing galaxies which contribute a total of $\sim20\%$ of
the projected mass inside the Einstein radius.  The baryonic fraction
is lower than the fraction $64\pm4$\% (18\% rms scatter) found by
\citet{gavazzi07} in averaging over 22 strong lenses, although is
within their spread of $40\% - 100\%$.  The halo alignment is
consistent with the baryons, both positionally and in orientation.
The elongation of both components is also consistent, although the
errors are relatively large, predominantly due to a degeneracy between
the halo elongation and the shear. We discuss the cluster shear
further in Section \ref{sec_cluster_shear}.

\begin{table}
\centering
\small
\begin{tabular}{cc}
\hline
Parameter & Minimised Value \\
\hline
$\rho_s$ & $(1.42\pm0.24)\times10^7\,{\rm M}_{\odot}{\rm kpc}^{-3}$\\
$\alpha$ & $1.42^{+0.24}_{-0.22}$ \\
$\Psi$ & $1.71^{+0.28}_{-0.38}$ M$_{\odot}$/L$_{B\odot}$ \\
$e_h$ & $1.12^{+0.09}_{-0.07}$ \\
$(\Delta x_h, \Delta y_h)$ & $(-0.03''\pm0.03'',0.06''\pm0.03'')$ \\
$\theta_h$ & $20.3^{\circ} \pm 3.5^{\circ}$ \\
$\gamma$ & $0.181 \pm 0.021 $ \\
$\theta_{\gamma}$ & $29.6^{\circ} \pm 4.1^{\circ}$ \\
$\Psi_{cg}$ & $3.6\pm1.9$ M$_{\odot}$/L$_{B\odot}$ \\
\hline
\end{tabular}
\normalsize
\caption{Best fit parameters for the dual component model (see Section
\ref{sec_model}).  1$\sigma$ errors are quoted.}
\label{tab_dual_params}
\end{table}

\begin{figure}
\epsfxsize=6.5cm
{\hfill
\epsfbox{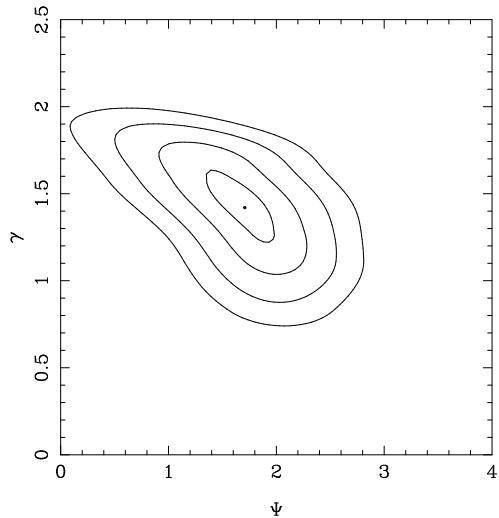}
\hfill}
\epsfverbosetrue
\caption{One-parameter confidence regions (1, 2, 3 \& 4$\sigma$) 
on the halo inner slope, $\gamma$, and M/L$_{B\odot}$ of the baryons,
$\Psi$, for the dual component model \citep[see][for
comparison]{treu04,dye05}. }
\label{chisq_conts}
\end{figure}

In Figure \ref{chisq_conts}, we plot the confidence levels on the
fitted inner halo slope and baryonic M/L, having marginalised over all
other parameters. The best fit slope is $\alpha=1.42^{+0.24}_{-0.22}$.
This is consistent with the value measured by \citet{treu04} of
$1.3^{+0.2}_{-0.4}$ with their baryons $+$ halo model, averaged over
three lenses although slightly higher than the value measured by
DW05, $\alpha=0.87^{+0.34}_{-0.27}$ (their 68\% CL) for the
Einstein ring 0047-2808.  The best fit M/L is $1.71^{+0.28}_{-0.38}$
M$_{\odot}$/L$_{B\odot}$ which compares to the local value for
ellipticals measured by \citet{gerhardt01} of $7.3\pm2.1$
M$_{\odot}$/L$_{B\odot}$.  To evolve this local measurement back to
the redshift of our lens, we use the findings of \citet{koopmans06}
that d$\log({\rm M/L}_{B})/$d$z=-0.69\pm0.08$. This gives a result of
M/L$_B=2.3\pm0.7$ M$_{\odot}$/L$_{B\odot}$, slightly higher than our
measurement but consistent within the errors. 

A source of error not included in Figure \ref{chisq_conts} is the
uncertainty on the halo scale radius $r_0$. However, the propagation
of this uncertainty is small. We find that a 10\% change in $r_0$
gives rise to a $\sim 1\%$ change in the minimised M/L and a
negligible change in the minimised slope. This is exactly the same
dependency found by DW05.

Finally, turning to the perturbing cluster galaxies, we find a best
fit $B$ band M/L of $\Psi_{cg}=3.6\pm1.9$
M$_{\odot}$/L$_{B\odot}$. This is in keeping with local ellipticals as
well as ellipticals found in $z\sim0.3$ clusters
\citep[e.g.,][]{natarajan04}. We note that if these cluster members
are omitted from the modelling then the best fit elongation of all
models (apart from the pure M/L) has to be significantly higher, $e_h
\simeq 1.4-1.5$ in order to fit the observed ring image. In this case,
the fit is not significantly degraded, but contradicts the concordant
view that isodensity contours have consistent or smaller elongations
than isophotal contours for a given system
\citep[e.g.,][]{koopmans06}.

\subsubsection{Cluster shear}
\label{sec_cluster_shear}

The best fit cluster shear predicted by the gNFW, power-law and SIE
models was found to be respectively ($\gamma=0.182\pm0.017$,
$\theta_{\gamma}=27.6^{\circ} \pm 3.3^{\circ}$),
($\gamma=0.180\pm0.018$, $\theta_{\gamma}=28.1^{\circ} \pm
3.4^{\circ}$) and ($\gamma=0.176\pm0.014$,
$\theta_{\gamma}=29.1^{\circ} \pm 2.9^{\circ}$).

As well as being self-consistent, this shear is in
agreement with what would be expected from an isothermal cluster mass
profile: Firstly, the tangential shear angle $\theta_{\gamma}$ is
consistent with the direction of the cluster centre at
$24^{\circ}$. Secondly, the magnitude is what would be expected given
the cluster proximity as the following approximate calculation
demonstrates: The ring lies $\sim1.8$ times further away from the
cluster centre than the outermost (and faintest) arc. Making the
reasonable assumption that this arc is an image of a $z\simgreat 2$
source and since the deflection angle varies very slowly with redshift
beyond $z\sim 2$, this gives a good approximation to the critical
radius for a source at infinity. Treating the cluster mass profile as
an isothermal sphere gives a predicted shear of $\gamma=0.28$ for a
$z\rightarrow \infty$ source at $r=1.8r_{\rm crit}$.  However, a
source at the lens redshift of $z=0.73$ experiences a shear of 0.63
times less (for our assumed cosmology) resulting in a shear of
$\gamma=0.175$.

The average of the shear from the gNFW, power-law SIE models was used
to distort the Sersic profile fitted to the observed galaxy
light. With this distorted profile, the dual component model gives a
best fit cluster shear of $\gamma=0.181 \pm 0.021$,
$\theta_{\gamma}=29.6^{\circ} \pm 4.1^{\circ}$.  The error budget here
includes the uncertainty on the distorted Sersic profile. This shear
agrees very well with the shear obtained from the gNFW, power-law and
SIE models, an indication that the lack of baryons in these models did
not bias the shear they predict.

The ability to predict the large scale cluster shear from the small
area of sky covered by the ring is quite remarkable. Although the
shear is fairly degenerate with the lens elongation, this degeneracy
is significantly higher in strong lens systems with only a point-like
source, to the extent that measurement of external shear in these
systems is impossible. This is a testament to the stronger constraints
provided by an extended source.

\subsection{Source reconstruction}
\label{sec_src_recon}

\begin{figure*}
\epsfxsize=16.5cm
{\hfill
\epsfbox{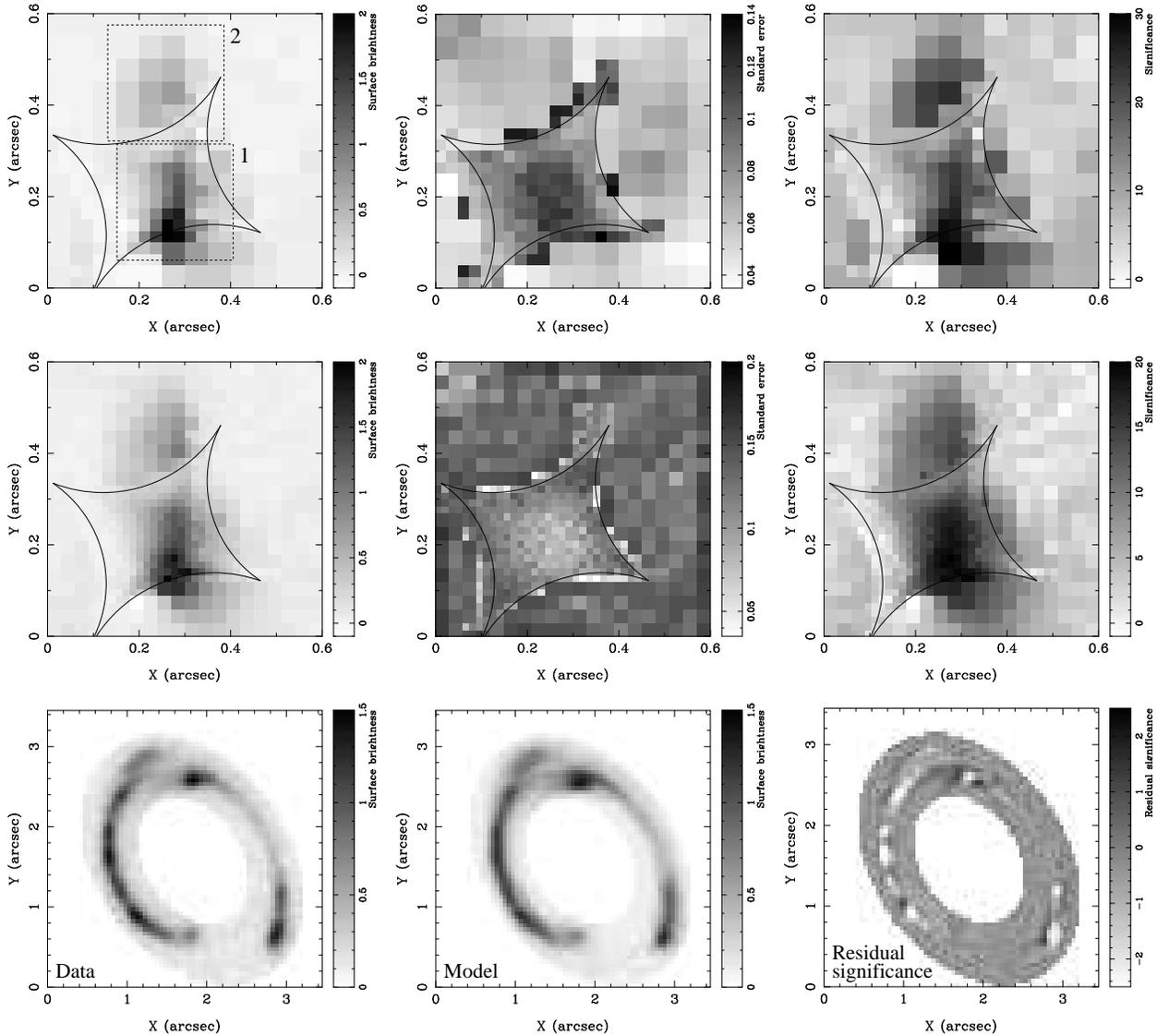}
\hfill}
\epsfverbosetrue
\caption{The reconstructed source from the best fit dual component
model. Reading from left to right along each row; {\em Top row} - the
unregularised reconstructed source, the standard errors map for the
reconstructed source and the significance map; {\em Middle row} - same
as top row but for the regularised source; {\em Bottom row} - the
masked observed ring, the image of the unregularised source and the
significance of the residuals. As can be seen, the lens model gives a
very good description of the observed ring morphology. The
reconstructed source shows two peaks. In the top left panel, the
dashed squares delineate the two source planes used in the dual-source
reconstruction (labeled 1,2). The source plane caustic is indicated
in the top two rows. In the source plane, $0.1''$ corresponds to 800pc.}
\label{recon_source_montage}
\end{figure*}

Figure \ref{recon_source_montage} shows the reconstructed source
obtained from the best fit dual component model.  The top row
corresponds to the unregularised solution and the middle row the
regularised case. In the bottom row we plot the observed image
alongside the image of the unregularised source and the significance
of the residuals. The residuals are within $\pm
2.5\sigma$. Considering some of the pixels in the observed image reach
a significance of $\sim 40\sigma$, this demonstrates the quality of
the fit.  Nevertheless, we have investigated the cause of these
residuals, i.e., whether they are due to an overly-coarse source
pixelisation, the parameterisation of the lens model or the PSF used
to smear images of lensed source pixels. The source pixelisation is
ruled out by the fact that the residuals are barely changed when we
use the image of the regularised source. We also repeated the
minimisation with different PSFs, including a PSF extracted from stars
in the field, but the best fit was obtained with the {\sl TinyTim}
model PSF (see Section \ref{sec_data}). We conclude therefore that the
residuals would probably be lessened by a small modification in the
parameterisation of the lens model. This would require an exhaustive
search through different parameterisations with only a small return and
no guarantee of finding a unique solution, hence we leave this for
possible future work.

\begin{figure}
\epsfxsize=7.5cm
{\hfill
\epsfbox{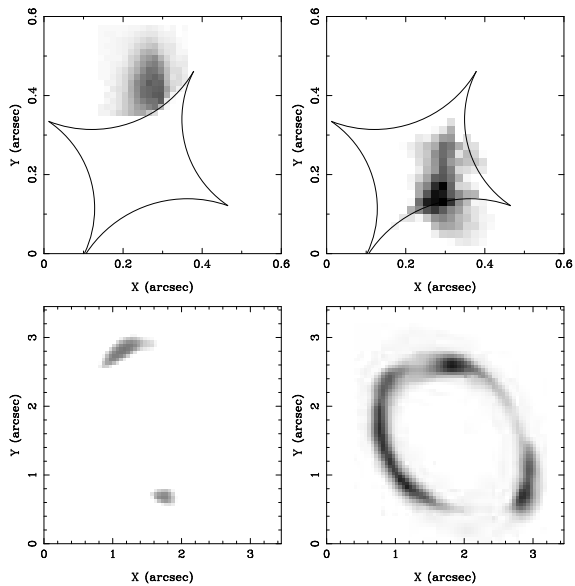}
\hfill}
\epsfverbosetrue
\caption{The contribution of each reconstructed source to the 
observed ring image. The top row shows the two sources and the bottom
row their respective images.}
\label{source_parts}
\end{figure}

The total magnification of the system (i.e., total ring flux divided
by total source flux) is $\sim 25$. There are clearly two peaks in the
source surface brightness distribution. To demonstrate the
contribution each peak makes to the observed Einstein ring image,
Figure \ref{source_parts} shows the image of each peak lensed
separately with the best fit dual component model. The dominant ring
structure is due to the brighter peak mostly contained within the
caustic.  The fainter peak lying nearer the top of the plane and
outside of the caustic is imaged into the westerly and easterly
extensions. Although not plotted, the double-peaked
nature of the reconstructed source obtained with the other models
is reproduced in every case.

\subsubsection{Dual source plane analysis}

We repeated the minimisation using the dual component model but with
two separate source planes as indicated in the top left panel of
Figure \ref{recon_source_montage}. Source plane 1 containing the
brighter source peak was held fixed at the redshift $z=3.07$ but the
redshift of source plane 2, $z_2$, was allowed to vary in the
minimisation.

Our results give a value of $z_2=3.13^{+0.73}_{-0.49}$ with an overall
fit that is not an improvement on the single source plane case. The
remaining minimised parameters of the model differ only negligibly to
those in Table \ref{tab_dual_params} for the single source plane case.
Although the errors are large, the redshift of the second source peak
is consistent with that of the brighter primary peak, suggesting that
these are two features within the same galaxy.

\section{Summary and Discussion}
\label{sec_summary}

Out of the five lens models assessed in this work, the Einstein ring
LBG J2135-0102 is best fit with our dual component model comprising a
gNFW dark matter halo that hosts a baryonic component following the
lens galaxy light. Of the remaining four single component models, the
gNFW, power-law and SIE models give an acceptable fit but the pure M/L
model is very strongly ruled out. Between the gNFW, power-law and SIE,
the gNFW gives a better fit with $\Delta \chi^2 = 1.2$ for six more
degrees of freedom compared to the power-law and $\Delta \chi^2 =
-5.4$ for two more degrees of freedom compared to the SIE. However,
all three models give a significantly worse fit than the dual component
model which has $\Delta \chi^2 = -17.4$ for six more degrees of
freedom compared to the gNFW model.

The dual component model predicts a projected baryonic contribution of
$(46\pm 11)$\% interior to the Einstein radius of $0.77''$. This is at
the low end (but consistent with) the spread in baryonic fraction of
40\% - 100\% measured by \citet{gavazzi07} over 22 strong lens
systems. We measure a M/L ratio of $1.71^{+0.28}_{-0.38}$
M$_{\odot}$/L$_{B\odot}$ for the baryonic component, in agreement with
the value for local ellipticals once the evolution measured by
\citet{koopmans06} is taken into consideration.

We have modelled the effect of the $z=0.33$ foreground cluster MACS
J2135.2-0102 on our lens solution, both in terms of the large scale
shear and convergence and also perturbations to the local potential
from nearby cluster members. The modelling accurately predicts the
cluster shear. In strong lens systems with only a point-like source,
this is impossible due to a strong degeneracy between the shear and
lens elongation.  Although a degeneracy exists between the cluster
shear and elongation in our modelling, the extra constraints provided
by the ring image greatly reduce it. This demonstrates the advantage
extended source systems give over point source systems.

In addition to shearing the ring image, the cluster shears the lens
galaxy light. Taking this shear into account, we find that the halo
and baryonic component are well aligned (i.e., in terms of the
centroid and orientation of the major axis) within uncertainties.
Failure to incorporate the local perturbing cluster members does not
significantly degrade the fit but results in a substantially higher
halo elongation ($e_h \simeq 1.4-1.5$) than that of the intrinsic lens
galaxy light. This perturbation is therefore required to prevent
contradiction with the accepted view that a given galaxy's
isophotes should be consistent with or more elongated than its
isodensity contours \citep[e.g.,][]{koopmans06}.

The best fit inner slope of the total mass is $\alpha=1.95\pm0.03$,
given by the gNFW model or $\alpha=2.09\pm0.04$ given by the power-law
model. This is in keeping with the findings of several strong lens
studies to date, for example \citet{koopmans06} who measured a slope
of $2.01^{+0.02}_{-0.03}$ averaged over 15 lenses and \citet{rusin03}
who measured a slope of $2.07\pm0.13$ averaged over 22 lenses. The
result lends further evidence towards the proposition that the slope
has little or no evolution out to $z\simeq 1$ \citep{koopmans06}.

With the dual component model, the best fit inner slope of the halo
component is $\alpha=1.42^{+0.24}_{-0.22}$.  This is consistent with
the value of $1.3^{+0.2}_{-0.4}$ averaged over three lenses by
\citet{treu04}, although slightly higher than the value
$\alpha=0.87^{+0.34}_{-0.27}$ measured for the $z=0.485$ lens
0047-2808 by DW05. To address whether this is consistent with pure CDM
simulations, the effect of halo contraction by the condensation of
baryons must be considered. \citet{blumenthal86} originally suggested
the model of adiabatic contraction.  More recently, \citet{gnedin04}
showed using high resolution CDM $+$ baryon simulations of halos that
the adiabatic contraction model over-predicts the increase in central
dark matter density \citep[see also][]{selwood05}. They provide
analytical fitting functions to describe how a NFW profile is
contracted by a collapsed baryonic profile of arbitrary inner
slope. The effective inner slope of the volume mass density profile
corresponding to our baryonic Sersic law is $\sim 1.7$ which,
according to these fitting functions, would contract a NFW profile
into a profile with a slope of $\sim 1.5$.  The slope determined in
our analysis for LBG\,J2135-0102 therefore corresponds to an
uncontracted slope of $\sim 1$, consistent with current CDM
simulations.

This paper presents detailed modelling of the inner halo mass profile
of only one Einstein ring system. To date, there are a further $\sim
25$ such systems imaged with the HST for which detailed dual component
models have not yet been fully published.  Analysis of these remaining
systems is crucial to provide a better measure of the mean and
dispersion of the inner halo slope thus greatly improving current
strong lensing constraints on the CDM model.

Finally, considering the reconstructed source, we can conclude that
the two peaks seen in the surface brightness distribution are most
likely structure within the same galaxy. The magnification of
the ring allows the source to be reconstructed with a
pixel scale of $\sim 1/2$ that of the observed image at the same
signal-to-noise and with approximately non-covariant pixels.  This
gain in spatial resolution means that integral field spectroscopy in
the optical and more importantly with laser adaptive optics in the
near infra-red offers the opportunity to spatially resolve the
star-formation, kinematic and chemical properties of individual H{\sc
ii} regions within this galaxy on scales of $\sim 200$pc \citep[for
example, see][]{swinbank06,swinbank07}.  This level of science is
already on a par with that which will be delivered by the next
generation of Extremely Large Telescopes.

\begin{flushleft}
{\bf Acknowledgements}
\end{flushleft}

SD is supported by PPARC.  IRS acknowledges support from the Royal
Society. AMS acknowledges support from PPARC. HE gratefully
acknowledges financial support from STScI grant HST-GO-10491. We thank
Johan Richard, Jean-Paul Kneib, Dan Stark, Richard Ellis, Graham Smith
and Chris Mullis for their work on defining the observational
properties of the LBG\,J213512.73-010143 system.  We thank an
anonymous referee for constructive comments which improved this work.

\end{document}